\documentclass[preprint,a4paper]{aastex63}

\usepackage{amsmath,amssymb}
\usepackage{natbib}
\usepackage{graphicx}
\usepackage{color}
\usepackage{url}
\usepackage{upgreek}

\received{July 19, 2019}
\accepted{December 16, 2019}
\submitjournal{Solar Physics}

\shorttitle{A Non-Linear Magnetic Field Calibration Method}
\shortauthors{J. Guo et al.}

\begin{document}

\title{A Non-Linear Magnetic Field Calibration Method for Filter-Based Magnetographs by Multilayer Perceptron}

\correspondingauthor{Kaifan Ji}
\email{jkf@ynao.ac.cn, guojingjing@bao.ac.cn}

\author{Jingjing Guo}
\affiliation{Key Laboratory of Solar Activity, National Astronomical Observatories, Chinese Academy of Sciences, Beijing 100101, China}
\affiliation{University of Chinese Academy of Sciences, Beijing 100049, China}

\author{Xianyong Bai}
\affiliation{Key Laboratory of Solar Activity, National Astronomical Observatories, Chinese Academy of Sciences, Beijing 100101, China}

\author{Yuanyong Deng}
\affiliation{Key Laboratory of Solar Activity, National Astronomical Observatories, Chinese Academy of Sciences, Beijing 100101, China}

\author{Hui Liu}
\affiliation{Yunnan Observatories, Chinese Academy of Sciences, Kunming 650216, China}

\author{Jiaben Lin}
\affiliation{Key Laboratory of Solar Activity, National Astronomical Observatories, Chinese Academy of Sciences, Beijing 100101, China}

\author{Jiangtao Su}
\affiliation{Key Laboratory of Solar Activity, National Astronomical Observatories, Chinese Academy of Sciences, Beijing 100101, China}

\author[0000-0003-1675-1995]{Xiao Yang}
\affiliation{Key Laboratory of Solar Activity, National Astronomical Observatories, Chinese Academy of Sciences, Beijing 100101, China}

\author{Kaifan Ji}
\affiliation{Yunnan Observatories, Chinese Academy of Sciences, Kunming 650216, China}

\begin{abstract}

For filter-based magnetographs, the linear calibration method under the weak-field assumption is usually adopted; this leads to magnetic saturation effect in the regions with strong magnetic field. This article explores a new method to overcome the above disadvantage using a multilayer perceptron network, which we call MagMLP, based on a back-propagation algorithm with one input layer, five hidden layers, and one output layer. We use the data from the \textit{Spectropolarimeter} (SP) on board \textit{Hinode} to simulate single-wavelength observations for the model training, and take into account the influence of the Doppler velocity field and the filling factor. The training results show that the linear fitting coefficient (LFC) of the transverse field reaches above 0.91, and that of the longitudinal field is above 0.98. The generalization of the models is good because the corresponding LFCs are above 0.9 for the test subsets. Compared with the linear calibration method, the MagMLP is much more effective on dealing with the magnetic saturation effect. Analyzing an active region, the results of the linear calibration present an evident magnetic saturation effect in the umbra regions; the corresponding systematic error reaches values greater than 1000 G in most areas, or even exceeds 2000 G at some pixels. However, the results of MagMLP at these locations are very close to the inversion results, and the systematic errors are basically within 300 G. In addition, we find that there are many ``bright spots'' and ``dark spots'' on the inclination angle images from the inversion results of \textit{Hinode}/SP with values of 180 and 0 degrees, respectively, where the inversion is not reliable and does not produce a good result; the MagMLP handles these points well.

\end{abstract}


\section{Introduction}
\label{sect:intro}

The solar magnetic field, which was first measured using the Zeeman effect \citep{hale1908probable}, is crucial to study the evolution, radiation, and nature of various solar activities events. Up to now, the instruments measuring solar magnetic fields can be classified as filter-based magnetographs (\textit{e.g.} the \textit{Narrowband Filter Imager} onboard \textit{Hinode} \citep{tsuneta2008solar} and the \textit{Solar Magnetic Field Telescope} (SMFT) installed at Huairou Solar Observing Station (HSOS) \citep{Guo1989SOLAR}), the spectral magnetographs such as the \textit{Spectropolarimeter} (SP) onboard \textit{Hinode} and the \textit{GREGOR Infrared Spectrograph} (GRIS) on the \textit{GERGOR} telescope installed at the Observatorio del Teide (OT) \citep{schmidt20121}, and the combination of the above two, for instance the \textit{Helioseismic and Magnetic Imager} onboard the \textit{Solar Dynamics Observatory} \citep{schou2012design} and the in-development two-dimensional real-time spectrograph \citep{deng2009progress}.

Generally speaking, the measurement of solar magnetic fields is indirect. The data we observe are the Stokes parameters $I(\lambda$), $Q(\lambda$), $U(\lambda$), and $V(\lambda$), where $\lambda$ is the wavelength. Vector magnetic fields and other atmospheric quantities on the solar surface are derived from the Stokes $I(\lambda$), $Q(\lambda$), $U(\lambda$), and $V(\lambda$) parameters. For spectral magnetographs with a large amount of spectral information, the spectral inversion method is generally adopted. The Levenberg-Marquardt algorithm was first developed and clearly explained in the seminal work by \citet{harvey1972line}. \citet{lagg2004retrieval} presented the so-called HeLIx code to explore the chromospheric diagnostic capabilities of the He \textsc{i} multiplet at 1083 nm. \citet{ramos2007bayesian} proposed using a `not-so-brute-force' procedure, a Markov chain Monte Carlo method, where marginal distributions of parameters can be obtained. An extremely interesting new generation of the inversion methods has been proposed by \citet{ramos2015sparse}.

In addition, the machine-learning (ML) method has been developed and seems to be an alternative and a complement to existing inversion methods. A principal component analysis (PCA) code was introduced by \citet{socas2001fast}. An artificial neural network (ANN) method was applied by \citet{carroll2001inversion}, and \citet{socas2003measuring}. In these cases, the system was trained with a set of synthetic Stokes profiles. For the purpose of handling huge amounts of data coming from future solar spectro-polarimeters much faster, statistical machine-learning techniques based on Mercer's kernel were applied to the inversion of photospheric magnetic fields from polarimetric data \citep{teng2015application}. Recently, a new inversion technique based on convolutional neural networks was designed by \citet{2019A&A...626A.102A}. It infers the thermodynamical and magnetic properties with a precision comparable to that of standard inversion techniques and is around one million times faster. More inversion techniques are described in the review written by \citet{del2016inversion}.

For filter-based magnetographs which take observations of the Stokes $I$, $Q$, $U$, and $V$ parameters just at one single-wavelength point, the linear calibration is usually adopted. The key step is to determine the calibration coefficients between the magnetic field and the Stokes parameters according to the weak-field assumption. In this case, the longitudinal magnetic field strength, $B_l$, and transverse magnetic field strength, $B_t$, can be derived by simplifying the radiation transfer equation under the weak-field assumption. They can be expressed using the calibration coefficient as
\begin{align}
 B_l &= C_l V,    \\
 B_t &= C_t (Q^2 + U^2)^{1/4},
\end{align}
where $C_l$ is the calibration coefficient of  $B_l$, and $C_t$ stands for that of  $B_t$. Five different algorithms have been applied to obtain the correlation coefficients \citep{bai2013calibration}. The method works well in the weak-field regions, while it dose not in the regions with strong magnetic field, such as the umbrae of sunspots where magnetic saturation occurs \citep{Bai2014Improved}. To overcome the saturation effect, a first-order polynomial approach was adopted to model the reversal of the polarization signal over the field strength by \citet{chae2007initial}.

However, the magnetic saturation effect is a non-linear phenomenon. The linear calibration cannot fix it well no matter what is the value of the coefficient. Due to the capability to approximate non-linear functions of artificial neural networks (ANN), in this article we attempt to use the multilayer perceptron (MLP) as an alternative to solve the magnetic saturation effect. The \textit{Full-disk MagnetoGraph} (FMG) \citep{2019RAA....19..157D} onboard the \textit{Advanced Space-based Solar Observatory} (ASO-S) \citep{2019RAA....19..156G}, which is scheduled to be launched in early 2022  will measure full-disk vector magnetic fields. It is a filter-based magnetograph working on Fe \textsc{i} 532.42 nm. The method studied in this can provide an alternative scheme for the magnetic field calibration at one wavelength and be useful for FMG and the existing filter-based magnetographs at HSOS.

The work presented here demonstrates the applicability of the MLP network to calibrate the magnetic field based on the Stokes $I$, $Q$, $U$, and $V$ observations at a single-wavelength point. The data used to train network models are from the \textit{Hinode}/SP. In Section~\ref{sec:data}, the data and how to preprocess them are introduced briefly. We provide the architecture and detailed training process of our network that we call MagMLP in Section~\ref{sec:method}. In Section~\ref{sec:result}, the training results using the MLP network are given with the analysis of the influence of some factors on the results. In Section~\ref{sec:comp}, the training results are compared with the linear calibration and the inversion results, demonstrating the capability of MagMLP to solve the ``bright spots'' and ``dark spots'' problem on the inversion results of the inclination angle. The conclusions are given in Section~\ref{sec:conc} with a discussion of our approach.

\section{Dataset and Data Preprocessing}
\label{sec:data}

\begin{figure}[!htbp]
	\centering
	\includegraphics[width=0.8\textwidth]{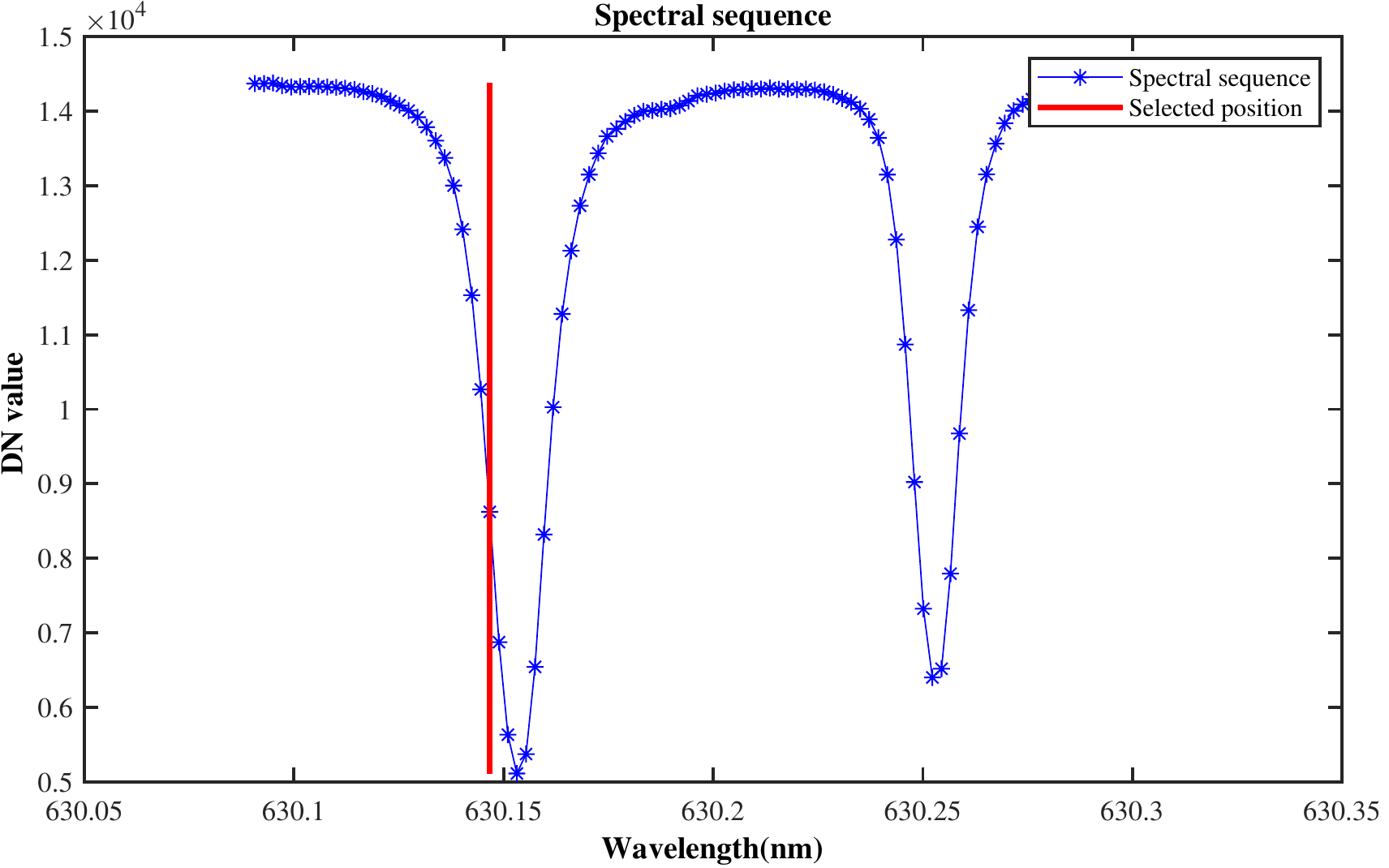}
    \caption{Schematic diagram for simulating the narrow-band observation using the actual observed data of \textit{Hinode}/SP.}
	\label{fig:1}
\end{figure}

For a training set, the completeness of the data is a basic requirement of neural networks; therefore, as many as possible active regions at different heliographic positions are included. Furthermore, in order to make the relationship between Stokes parameters and magnetic parameters consistent, the data are selected with the same resolution and exposure time. The \textit{Hinode}/SP data are suitable for our study. Level 1 and Level 2 data from the \textit{Hinode}/SP, were selected as datasets and downloaded from \url{http://csac.hao.ucar.edu/sp_data.php}. The training set is composed of a total of 139 frames of solar active region data, 98 frames of which correspond to 2014 from 10:17 UT-26 July to 22:30 UT-30 September and from 21:22 UT-4 November to 00:30 UT-20 December, 14 frames in 2015 from 00:00 UT-1 January to 09:11 UT-16 February, and 27 frames in 2017 from 08:26 UT-22 August to 19:46 UT-29 September. The test set consists of 37 frames of active regions: 9 frames in 2011 from 14:23 UT-28 July to 12:10 UT-7 November, 16 frames in 2014 from 12:21 UT-4 January to 12:34 UT-18 April and from 15:40 UT-21 October to 23:41 UT-24 October, 5 frames in 2018 from 13:25 UT-6 February to 05:28 UT-21 June, and 7 frames in 2019 from 13:57 UT-21 March to 06:00 UT-8 May.  Level 1 data are the calibrated 3D data (spectral \textit{per} spatial \textit{per} four Stokes parameters) ready for scientific analysis. Level 2 data are the outputs from Stokes inversion using the High Altitude Observatory (HAO) Merlin inversion code developed under the Community Spectra-polarimetric Analysis Center. Each Level 2 dataset contains 42 extensions, one \textit{per} inversion parameter or ancillary data product. More details about the data can be found in \url{http://www2.hao.ucar.edu/csac/csac-data/sp-data-description}. For the sake of simulating observations of filter-type magnetographs with one wavelength point, the Stokes $I$, $Q$, $U$, and $V$ at the wavelength position at $-0.063$ \AA\ from the center of Fe \textsc{i} 6301\AA\ were selected, as shown in Figure~\ref{fig:1}.

Theoretically, the Stokes $I$, $Q$, $U$, and $V$ are used to calibrate the vector magnetic field directly. In actual observations, the Doppler velocity field, \textit{Vel}, will be coupled in the polarization measurements at a single-wavelength point because the \textit{Vel} will cause the working line shift, which means that the Stokes $I$, $Q$, $U$, and $V$ will be observed at slightly different wavelength points. In order to eliminate the influence of the \textit{Vel}, we also consider adding the \textit{Vel} to the input parameters for comparison.

In the linear calibration process, the transverse field strength, $B_t$, and longitudinal field strength, $B_l$, cannot be obtained directly, while the training data from \textit{Hinode}/SP provide merely the magnetic field strength, $B$, and inclination angle, $\varphi$. The relationship among them can be expressed as
\begin{gather}
B_l = B\cos(\varphi),    \\
B_t = B\sin(\varphi),    \\
B = \sqrt{B_l^2 + B_t^2}.
\end{gather}
Therefore, we consider using $(B_t, B_l)$ or $(B, \varphi)$ as the parameters of the output layer.

In addition, we consider the influence of the filling factor, $\alpha$, which is used to treat spatially unresolved magnetic fields. The relationship between the magnetic field strength and  $\alpha$ can be written as
\begin{equation}
B = \alpha B',
\end{equation}
where $B$ is the magnetic field strength without taking into account $\alpha$, and $B'$ denotes the magnetic field strength taking $\alpha$ into account. $B'$ and $\alpha$ can be found from the \textit{Hinode}/SP Level 2 data because $\alpha$ is considered during the inversion process. The longitudinal (transverse) magnetic filed strength $ B'_l$ ($B'_t$) taking into account $\alpha$ can be written as
\begin{align}
B'_l &= B' \cos(\varphi),    \\
B'_t &= B' \sin(\varphi).
\end{align}

In conclusion, the magnetic field, $B'$, inclination angle, $\varphi$, Doppler velocity field, \textit{Vel}, and filling factor, $\alpha$, are extracted from the corresponding Level 2 fits files, while $B$, $B_l$, $B_t$, $B'_l$, and $B'_t$ can be calculated from the above four measurements.

In order to overcome the influence of noise in the Level 1 data and balance the coverage of umbras, penumbras, and quiet regions, we take the following steps to do the data preprocessing.
	
(i) Select regions of interests. Maintain as many as possible selected rectangular areas with strong magnetic field, and avoid including quiet regions and the solar limb.
	
(ii) Eliminate the noises on input and output data. Eliminate impulsive noise by median filtering using a $5\times 5$ operator, remove the pixels with magnetic field lower than 200 G, and leave out the single speckles for which the set of 8-adjacency connected components is less than 20 pixels.
	
(iii) There are some differences among the images. According to the relationship among the Stokes parameters, each image has been equalized to eliminate such differences. The Stokes $I$ (total light intensity) is normalized via dividing by the median of the quiet region in the $I$ image itself. Dividing by $I$, the Stokes $Q$, $U$, and $V$ are normalized. Thus the normalized Stokes parameters are
\begin{equation}
I_{\mathrm{norm}} = \frac{I}{I_{\mathrm{median}}},\quad Q_{\mathrm{norm}} = \frac{Q}{I},\quad U_{\mathrm{norm}}=\frac{U}{I},\quad V_{\mathrm{norm}}=\frac{V}{I}.
\end{equation}

(iv) Equalize the data. In terms of plotting the histograms for the magnetic field strength, inclination angle, transverse field strength, and longitudinal field strength, a fixed number of pixels are selected from evenly divided bins if the bin has pixels more than the fixed number, otherwise all pixels are taken. Each selected pixel is taken as a sample in the training set. In total, there are $\approx 12.9 \times 10^6$ samples in the training set. The strategy is that all the pixels are divided into equal-interval bins with a fixed interval, $F_c$, then a fixed number, $N$, of pixels is randomly selected from every bin if its total number of pixels is larger than $N$, otherwise, all pixels are taken as samples.
	
\begin{figure}[!htbp]
	\centering
	\includegraphics[width=\textwidth]{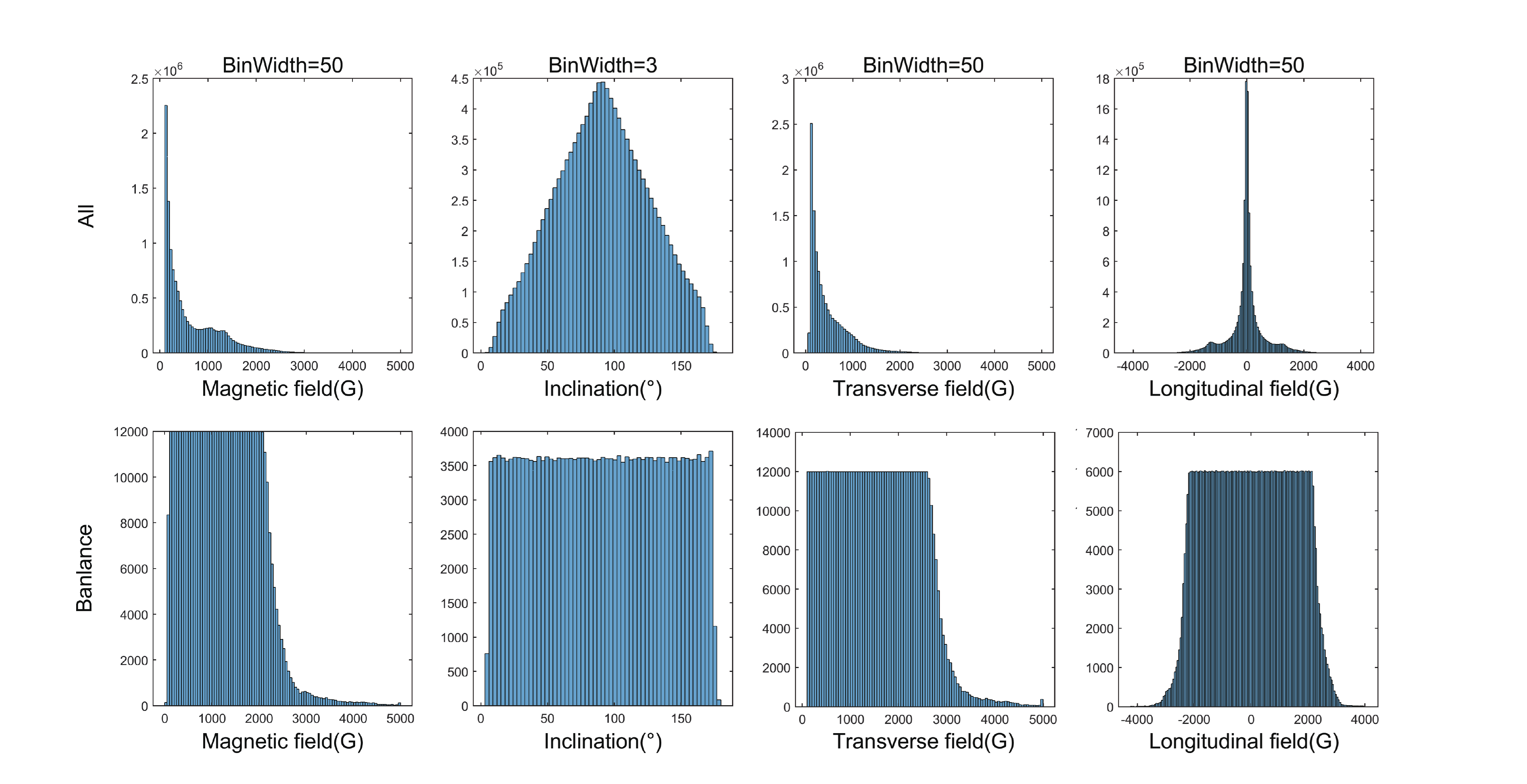}
    \caption{Histograms comparing the magnetic field, inclination angle, transverse field, and longitudinal field before and after equalization. The \textit{top row} shows the total data, and the \textit{bottom row} shows the equalized sample.}
    \label{fig:2}
\end{figure}

As an example of the procedure, for the total magnetic field strength, the pixels are divided into equal-interval bins with $F_c = 5$ G, and with $N = 1000$ in each bin. Then, there are $\approx 675 \times 10^3$ samples randomly obtained from all. According to the similar principle, there are about $203\times 10^3$, $558\times 10^3$, and $586\times 10^3$ samples selected for the inclination angle, transverse field, and longitudinal field with $F_c$ of 1 degree, 5 G, and 5 G, and $N$ of 1000, 1000, and 500, respectively. In Figure~\ref{fig:2}, the four panels on the top row show the distributions of all the pixels, while the bottom row shows the distributions of the equalized data.

(v) Balance the weight of the data. The input parameters are normalized from the original $[\min,\max]$ to $ [-1,1]$ before the training, so that they are moderately weighted.

\section{Method}
\label{sec:method}

At present, there is no theory that can produce accurately the magnetic field parameters from a single-wavelength point observed by a filter-based magnetograph as the multi-wavelength points inversion does. The linear calibration under the weak-field assumption simplifies the non-linear relationship between the Stokes parameters and the magnetic field. A multilayer perceptron trained by a back-propagation (BP) algorithm can be regarded as a practical tool for implementing non-linear input--output mapping of a general nature \citep{haykin2009neural}. For solving the problem, we try to build the relationship using a MLP which does not rely on a specific mathematical function. Essentially, it is a multiple non-linear regression problem to employ the results of Stokes inversion by simulating a single-wavelength observation to conduct a magnetic field calibration. Multiple non-linear regressions could be done by single or multilayer neural networks that can flexibly set the number of input variables. In this article, the magnetic field calibration is based on a BP multilayer fully connected neural network, \textit{i.e.} the MLP. The BP algorithm was originally introduced in the 1970s, but its importance was not fully appreciated until a famous paper by \citet{Rumelhart1986Learning}.

\begin{figure}[!htbp]
	\centering
	\includegraphics[width=0.8\textwidth]{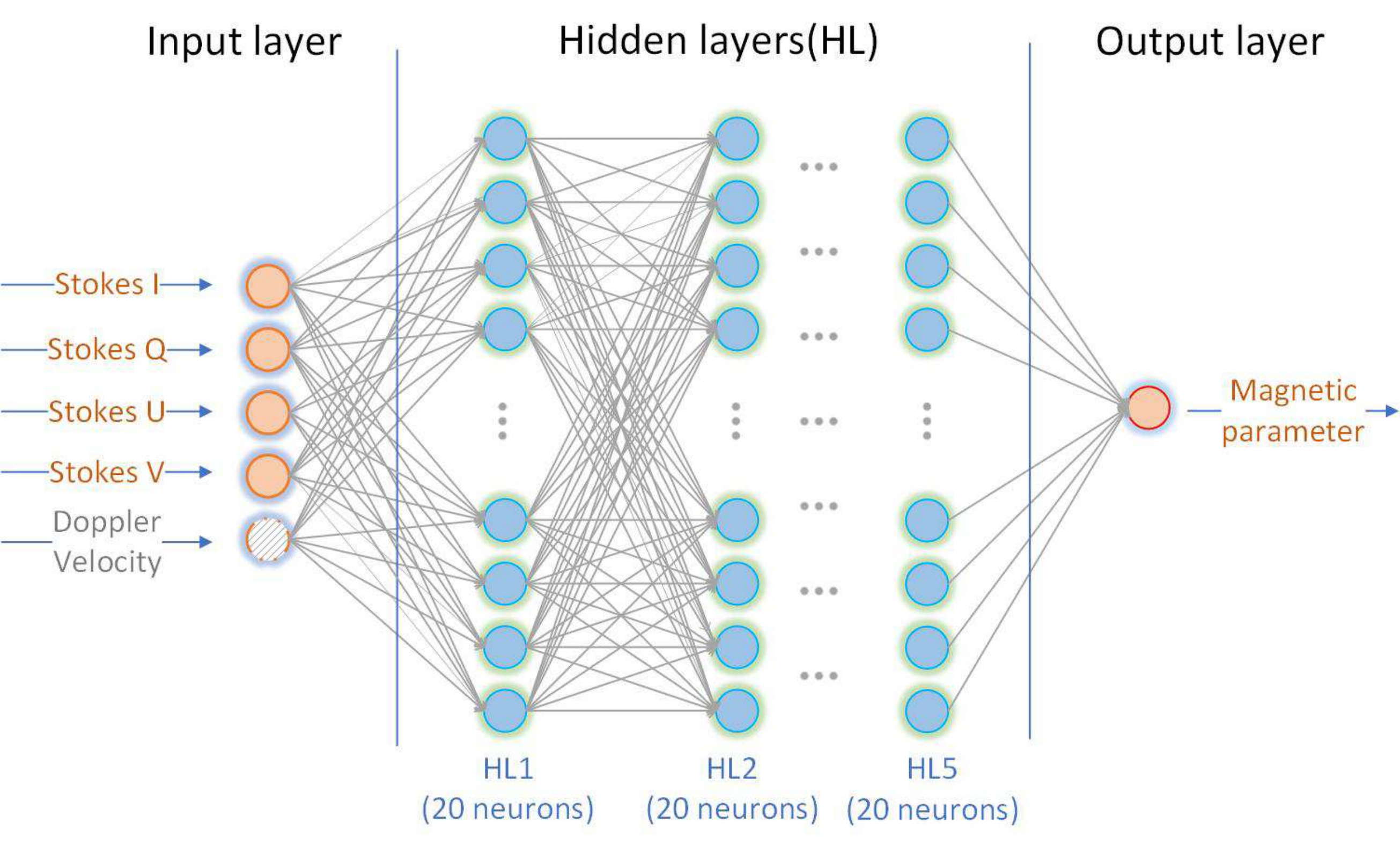}
	\caption{Schematic architecture of the MLP network used in this article.}
	\label{fig:3}
\end{figure}

The gradient BP algorithm is the essential ``learning engine'' powering most of this work. A number of BP network models are constructed to conduct the trial-and-error experiments. If multiple parameters are designed in the output layer, the network structure will become complex. For simplicity, one parameter in the output layer is used to train and we find a satisfactory network model, which we called MagMLP. It is made of one input layer, five hidden layers, in each of which 20 neurons are employed, and one output layer. Each hidden layer is a fully connected one. Figure~\ref{fig:3} shows the architecture.

The Bayesian regularization algorithm \citep{MacKayBayesian, Foresee1997Gauss}, which updates the weight and bias values according to the Levenberg-Marquardt optimization, is employed to get well generalized neural networks. It determines the correct combination of squared errors and weights to make networks generalize well.

\begin{table}
	\caption{Main training parameters with their values according to the Bayesian regularization.} 	\label{tab:1}
	\centering
	\begin{tabular}{l l l}
		\hline\hline
		Parameters & Value & Content  \\
		\hline
		epochs     		& 1500  & Maximum number of epochs to train \\
		$\mu$        	& 0.005 & Marquardt adjustment parameter \\
		$\mu\_$dec    	& 0.1   & Decrease factor for $\mu$  \\
		$\mu\_$inc      & 10 	& Increase factor for $\mu$  \\
		$\mu\_$max     	& $ 1 \times e^{10} $   & Maximum value for $\mu$  \\	
		max$\_$fail   	& 20    & Maximum validation failures \\
		min$\_$grad   	& 0.001 & Minimum performance gradient \\
		\hline
	\end{tabular}
\end{table}

Training occurs according to the Bayesian regularization, and the main training parameters are shown with their values in Table~\ref{tab:1}. We use the mean squares error (MSE) as the loss function, and the sigmoid function as the activation function in all hidden layers. The sigmoid function is sometimes referred to as logistic regression or in a shorter way as ``logsig''; its expression is
\begin{equation}
\mathrm{logsig}(x) = 1 / (1 + \mathrm{e}^{-x}).
\end{equation}
The linear transfer function, $\mathrm{output}(x) = x$, is used in the output layer. The whole training process was carried out into five stages. In each stage the value of the learning rate and the number of training samples adopted different settings in order to make the network converge faster and more efficiently. The learning rate is set using an exponential decay function
\begin{equation}
\mathrm{lr} = 0.3^{n-1} \mathrm{lr}_0,
\end{equation}
where $n$ is the number of the stage, and $\mathrm{lr}_0$ is the initial learning rate set as 0.03. The number of the training samples increases with the number of the stage. First, a small number of samples are used to shorten the training time, and then much more samples are adopted in the following training process. For each stage, data equalization, as described in the previous section, is carried out again to randomly select samples where $F_c$ has the same value but $N$ increases successively by 50, 100, 500, 800, and 1000. Then repeat the fifth stage several times to go on training until obtain the approving models. Training stops when any of these conditions occur:
		
(i) The maximum number of epochs (repetitions) is reached.

(ii) The maximum amount of time is exceeded.	

(iii) Performance is minimized to the goal.	

(iv) The performance gradient falls below min$\_$grad.	

(v) $\mu$ exceeds $\mu\_$max.

(vi) Validation performance does not increase more than max$\_$fail times since the last time it increased.
	
\begin{figure}[!htbp]
	\centering	
	\includegraphics[width=0.6\textwidth]{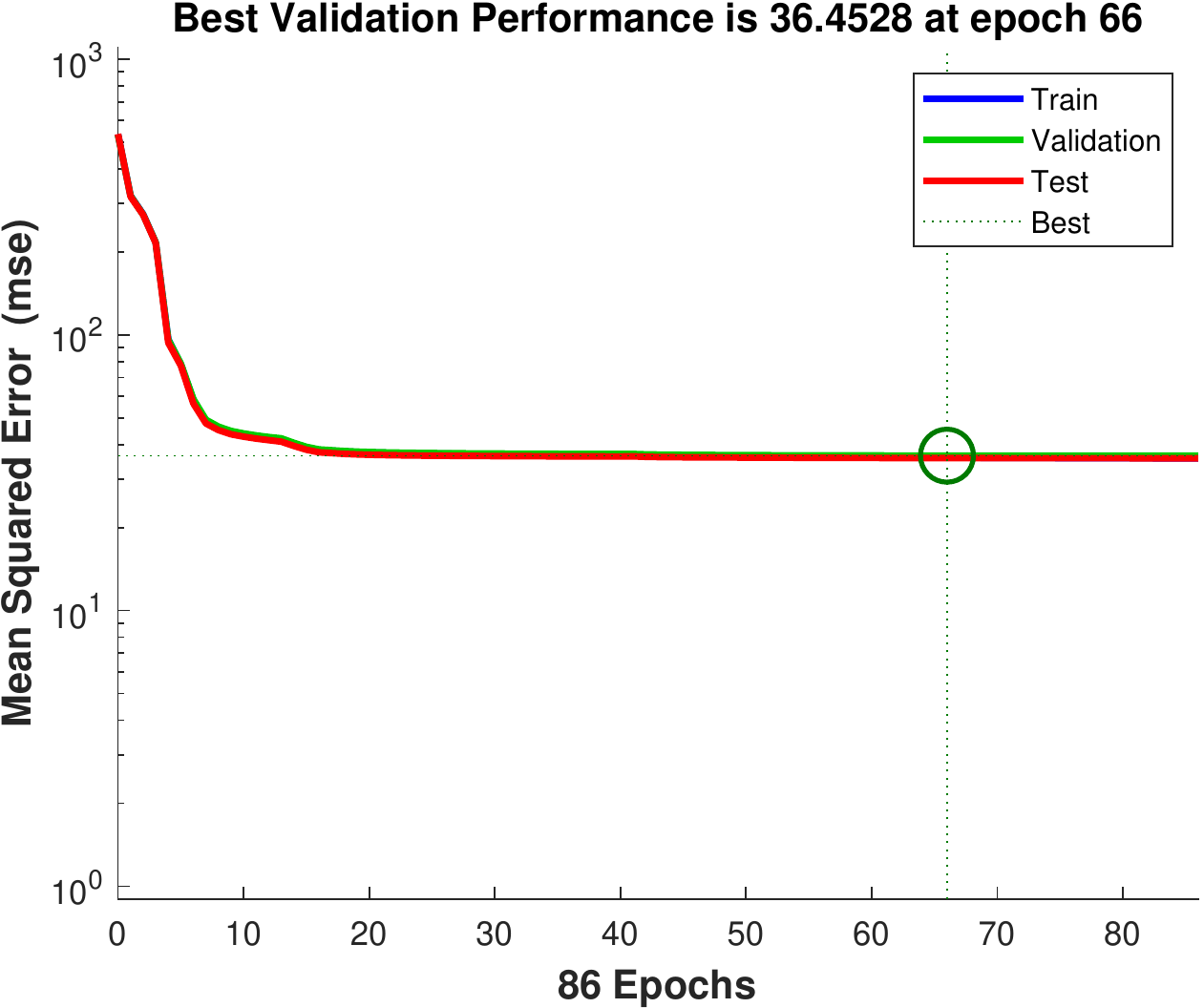}
    \caption{Loss function of the inclination model in the first stage of training with $I$, $Q$, $U$, $V$, and \textit{Vel} as input parameters.}
    \label{fig:4}
\end{figure}

In our training process, the values of the parameters related to conditions (ii), (iii), and (iv) are set sufficiently large or small so that the training will not terminate because of these three conditions. Condition (i) happens occasionally in the top three stages of the training because the value is relatively large, and condition (v) occurs occasionally during the training. The final network models used in this article are all stopped by condition (vi). Taking the inclination model in Figure~\ref{fig:4} as an example, the model training ends at the 86 epoch, and the final model is completed at the epoch 66.

The calculation of network models is done in parallel cores making a parallel pool open. The computer has two Intel Xeon E5-2609 CPUs with a processor base frequency of 1.7GHz and 8 cores for each CPU, and a 32 GB memory. The calculation connects 12 cores and an epoch takes about 56 seconds when the size of sample is about $670\times 10^3$. The MagMLP network models used in this work and the corresponding data and code can be downloaded from our repository \url{https://github.com/Guo-JJ/MagMLP}.
	
In terms of data preprocessing, the weight balance is carried out to make the MagMLP more conductive to learn and mine the inherent relationship between the Stokes parameters and magnetic parameters. We also analyzed the effect of several parameters on the magnetic field and their characteristics, given that we ultimately built 12 MagMLPs having different input layer or output layer as described in Table~\ref{tab:2} to conduct model training until they all converged. Then we verified the effectiveness of MagMLP by analyzing the linear fitting coefficients (LFC) and residual errors between the training results and the targets on the training set and comparing the results on the test set with the data from the Stokes inversion.

\begin{table}
	\caption{The linear fitting coefficients (LFC) of training results for 12 network models.}
	\label{tab:2}
	\centering
	\begin{tabular}{c c c c c}
		\hline\hline
		\multicolumn{1}{c}{Output}
		& \multicolumn{2}{c}{$ I,Q,U,V$,\textit{Vel}} &  \multicolumn{2}{c}{$ I,Q,U,V $}  \\
		\cline{2-3} \cline{4-5}
		\multicolumn{1}{c}{parameter}
		&\multicolumn{1}{c}{Train subset }  & {Test subset }
		&\multicolumn{1}{c}{Train subset }  & {Test subset }  \\
		\hline
		$ B{'} $       & 0.94396 & 0.90205 & 0.92643 & 0.90985   \\
		$ \varphi $    & 0.98841 & 1.00090 & 0.98799 & 1.00350   \\
		$ B_{t}{'}  $  & 0.93385 & 0.97219 & 0.92007 & 0.99064   \\
		$ B_{l}{'} $   & 0.98550 & 1.04630 & 0.98160 & 1.05370   \\
		$ B_{t}  $     & 0.94478 & 0.98967 & 0.92945 & 1.01040   \\
		$ B_{l} $      & 0.98503 & 1.03180 & 0.98286 & 1.04490   \\
		\hline
	\end{tabular}
\end{table}

\section{Results}
\label{sec:result}

\subsection{Training Results}

\begin{figure}[!htbp]
	\centering	
	\includegraphics[width=0.5\textwidth]{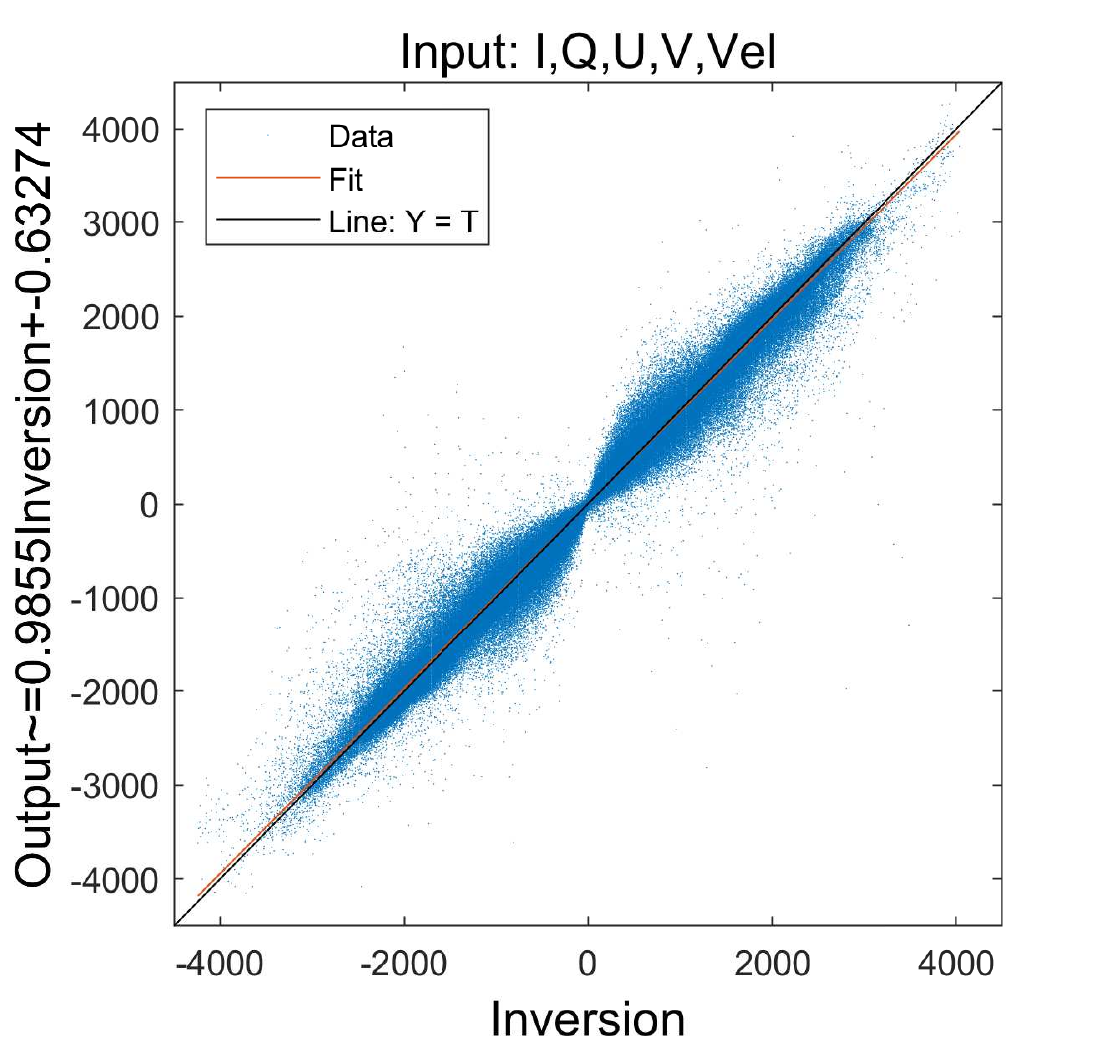}
    \caption{The scatter diagram displaying the distribution relationship between the training results and the inversion data for the $B'_l$.}
    \label{fig:5}
\end{figure}

The MagMLP is a supervised regression network model. Ideally, the best outcome would be that its output value, $Y$, should be one-to-one equal to the target value, $T$, as $Y = T$ with the LFC, $k = 1$. In fact, $Y$ can be infinitely close to and cannot equal $T$, and we use the line with slope $k$ as a fitting between $Y$ and $T$ to represent the performance of MagMLP. After having the models trained, we input all the training set to make a prediction and carry out a line fitting between the output data, $Y$, and inversion data, $T$, for every model. We take $B'_l$ as an example to illustrate this, as shown in Figure~\ref{fig:5}. The LFC of the training set is 0.9855 close to the ideal LFC 1. The data points are distributed centrally around the straight line of $Y = T$ and cannot achieve a perfect straight line distribution. This is due to the dual effects of network model systematic errors and Stokes inversion errors. Since the errors are unavoidable, they are neglected in the performance evaluation.

Considering the influence of the Doppler velocity, \textit{Vel}, and the filling factor, $\alpha$, 12 network models are built corresponding to six magnetic field measurements, $B'$, $\varphi$, $B'_t$, $B'_l$, $B_t$, $B_l$, as shown in Table~\ref{tab:2}. The input layer on half of the training models includes \textit{Vel}, and the other six models use $I$, $Q$, $U$, and $V$ as input without \textit{Vel}. The LFCs of every model are listed in Table~\ref{tab:2} on training subsets from the last training model and on the test set dropping the pixel points below 200 G.

The LFCs of the training subset are all above 0.9, and the results of containing the \textit{Vel} to train are all more or less higher than those without the \textit{Vel}. This indicates that the MagMLP effectively works on the magnetic field calibration for filter-based magnetographs, and adding \textit{Vel} to train reduces the complexity of the problem and improves the accuracy. From the above analysis, it is found that the network performance with the \textit{Vel} is better, and the existence of the Doppler velocity increases the complexity of the magnetic field calibration. The data used in this article all adopt the training results with \textit{Vel} to make the performance of the network better. On the other hand, the LFCs of $B_t$ and $B_l$ models without considering $\alpha$ are higher than those considering $\alpha$. On this point, further analysis and discussion will be made.

In order to further verify the generalization ability of the models, about $9.3\times 10^6$ sample data are obtained after all the data in the test set are preprocessed. The LFCs of all the models on the test subset are calculated with about $ 9.3\times 10^6$ sample data, and the results are shown in the third and fifth columns of Table~\ref{tab:2}. The LFCs of all models are above 0.9, very close to 1, except that the LFCs of the two models of $B$ are slightly lower. Therefore, it can be inferred that the models have a good generalization ability.

\subsection{Verification of the Network Availability }

\begin{figure}[!htbp]
	\centering
	\begin{minipage}{\textwidth}
		\centering
		\includegraphics[width=\textwidth]{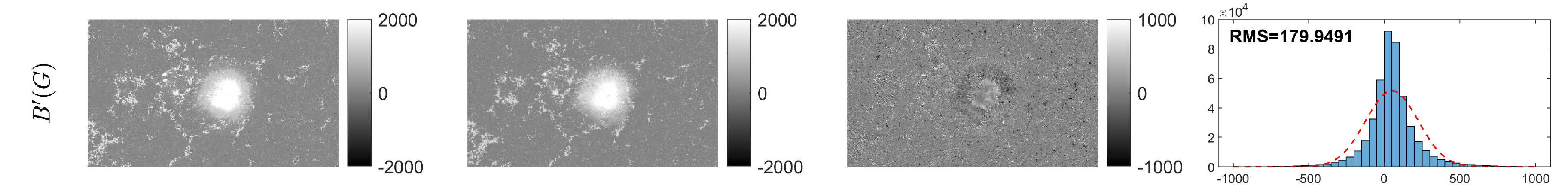}
	\end{minipage}
	
	\begin{minipage}{\textwidth}
		\centering
		\includegraphics[width=\textwidth]{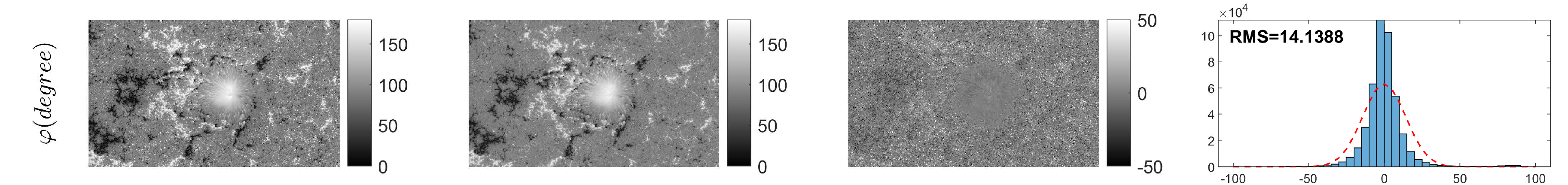}
	\end{minipage}
	
	\begin{minipage}{\textwidth}
		\centering
		\includegraphics[width=\textwidth]{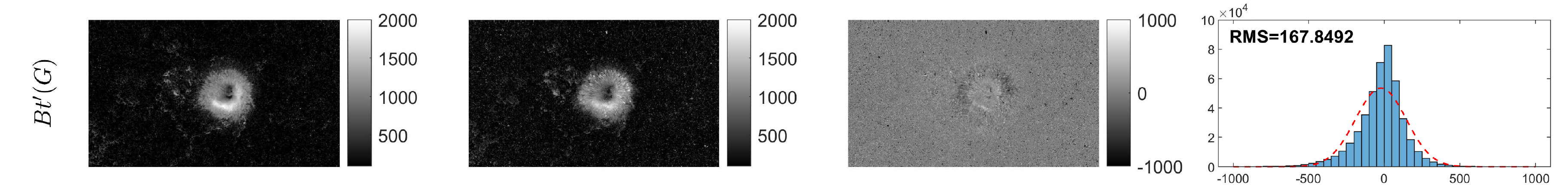}
	\end{minipage}	
	
	\begin{minipage}{\textwidth}
		\centering
		\includegraphics[width=\textwidth]{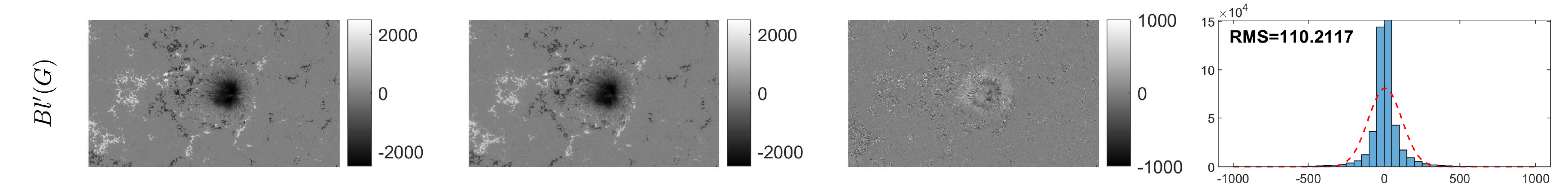}
	\end{minipage}
	
	\begin{minipage}{\textwidth}
		\centering
		\includegraphics[width=\textwidth]{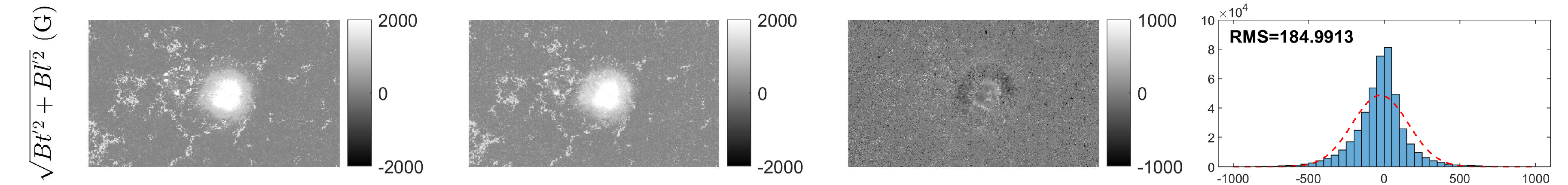}
	\end{minipage}

    \caption{The results predicted on one of the test sets by MagMLP. The \textit{leftmost colomn} is the target data from the magnetic field inversion. The \textit{second column} displays the results of the networks prediction. The \textit{third column} shows the residual of target data with the prediction. The \textit{rightmost column} is the histogram of the errors. The \textit{upper four rows} show the results for $B'$, $\varphi$, $B'_t$, and $B'_l$ from top to bottom. The results of the combination of transverse field and horizontal field, $B''$, using Eq. (5) is shown in the \textit{bottom row}.}
    \label{fig:6}	
\end{figure}

We select the active region (AR) 12738 observed with \textit{Hinode}/SP at 14:13 UT on April 20 2019 to test the models. The outputs of the models for $B'$, $B'_t$, $B'_l$, and $\varphi$ are shown in Figure~\ref{fig:6}. The upper four rows show the results for $B'$, $\varphi$, $B'_t$, and $B'_l$ from top to bottom. The leftmost is the target data from the Stokes inversion. The second column displays the results of the network prediction. The third column shows the residual errors of the target data with the prediction. The rightmost is the histogram of the residual errors.

Even though the training results are not all small-scale, details exactly being the same as with the target data, most of them approach the target data morphologically and numerically. The residual diagrams in the third column of Figure~\ref{fig:6} are gray in most areas. The residual errors of $\varphi$ are distributed mostly within $\pm 30$ degrees, and for the other parameters, a majority of the residual errors are within $\pm 300$ G. The residual histograms in the rightmost column of Figure~\ref{fig:6} show the results more intuitively; their distribution does not conform to the normal distribution displayed as a red dotted curve, error data close to zero are more densely distributed and are higher than the normal distribution curve. The root mean square (RMS) of the residual errors, which is an effective value of the residual errors, is used to analyze the results predicted here. The RMS for $B'$, $\varphi$, $B'_t$, and $B'_l$ are about 179.9 G, 14.1 degrees, 167.8 G, 110.2 G, respectively.

To understand the pros and cons between the separate training of $B'_t$ and $B'_l$ and the direct training of $B'$, the results of the separate training of $B'_t$ and $B'_l$ are combined to calculate $B''$. The results of $B''$ using Equation (5) are shown on the bottom row of Figure~\ref{fig:6}. The small-scale details are similar to the target data. The residual histograms of $B'$ and $B''$ are distributed symmetrically and have similar distributions. In addition, the RMS of the residual errors for $B'$ is 179.9491 G, about 5 G lower than $B''$. From the above analysis, the results of the two training strategies are very close, and it is hard to say which one is better. The strategy of separating the training $B_t$ and $B_l$ is adopted for the comparison with linear calibration in the following.

\subsection{Influence of the Filling Factor}

\begin{figure}[!htbp]
	\centering
	\begin{minipage}{\textwidth}
		\centering
		\includegraphics[width=\textwidth]{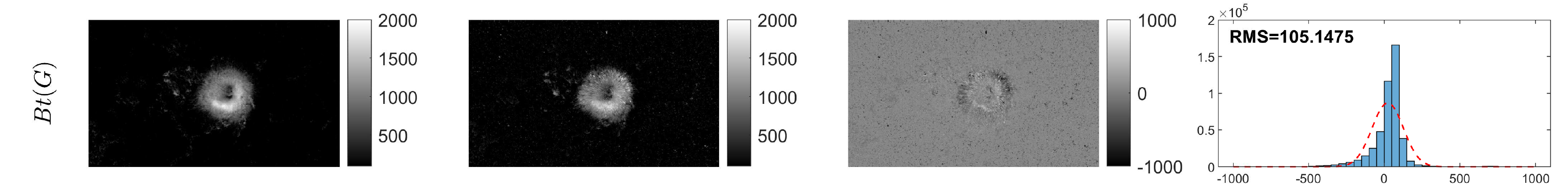}
	\end{minipage}
	
	\begin{minipage}{\textwidth}
		\centering
		\includegraphics[width=\textwidth]{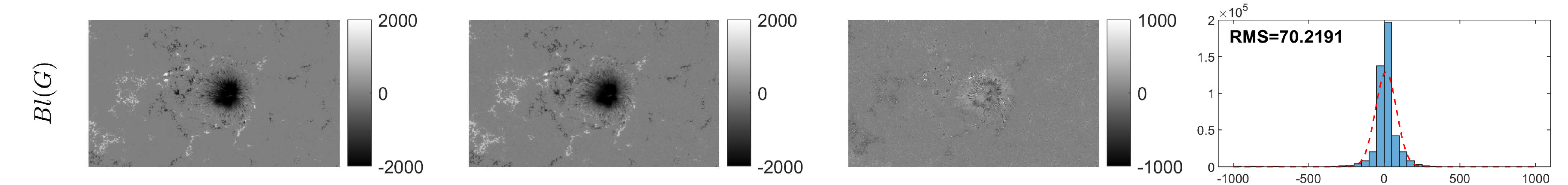}
	\end{minipage}		
	
    \caption{The training results of $B_t$ and $B_l$. The \textit{top row} corresponds to $B_t$, the \textit{bottom} to $B_l$. The \textit{panels} have similar information as those in Figure~\ref{fig:6}, from left to right.}
    \label{fig:7}
\end{figure}

Following the above comparative analysis, the network models are established for $B_t$ and $B_l$ without taking into account the filling factor $\alpha$. Considering $\alpha$ in Figure~\ref{fig:7}, which contains the same information as Figure~\ref{fig:6} from left to right, shows the training results for $B_t$ and $B_l$. Viewing the residual diagrams, the gray region covers more spaces in the training results without $\alpha$ than those considering $\alpha$ for both $B_t$ and $B_l$. From the residual histogram, the network models of $B_t$ and $B_l$ have a better convergence than in those of $B'_t$ and $B'_l$. For $B_t$, the RMS of residual errors is 105.1475 G, which is about 62 G lower than $B'_t$ (as the third row shows in Figure~\ref{fig:6}). For $B_l$, it is 70.2191 G, which is about 40 G lower than $B'_l$ (as the fourth row shows in Figure~\ref{fig:6}). The relations between the magnetic field quantities and the Stokes parameters become complicated when taking $\alpha$ into account. This indicates that the lower the complexity of the problem, the better the models convergence, that is, the models without considering the filling factor get significantly better results than those taking into account $\alpha$.

\section{Comparison and Analysis of the Results}
\label{sec:comp}

According to the above analysis, both the Doppler velocity field and $\alpha$ have influence on the training results. The effect of $\alpha$ is relatively distinct, so the addition of $\alpha$ makes the regression model more complicated. For filter-based magnetographs, the magnetic saturation is evident in magnetic field calibrated by the linear method, which brings a large systematic error into the absolute magnetic filed strength. In addition, we found that there are many ``bright spots'' on the results of the inclination angle from the Stokes inversion. By communicating with the staff of \textit{Hinode}/SP, we were informed that these pixels appear to be places where the inversion did not manage to find a good solution. In most of these pixels the flux density is quite low, indicating that there is not a high signal. The two aspects are discussed in the following section comparing the analysis with the training results of MagMLP.

\subsection{Compare the Results of MagMLP with a Linear Calibration }

\begin{figure}[!htbp]
	\centering
	\includegraphics[width=\textwidth]{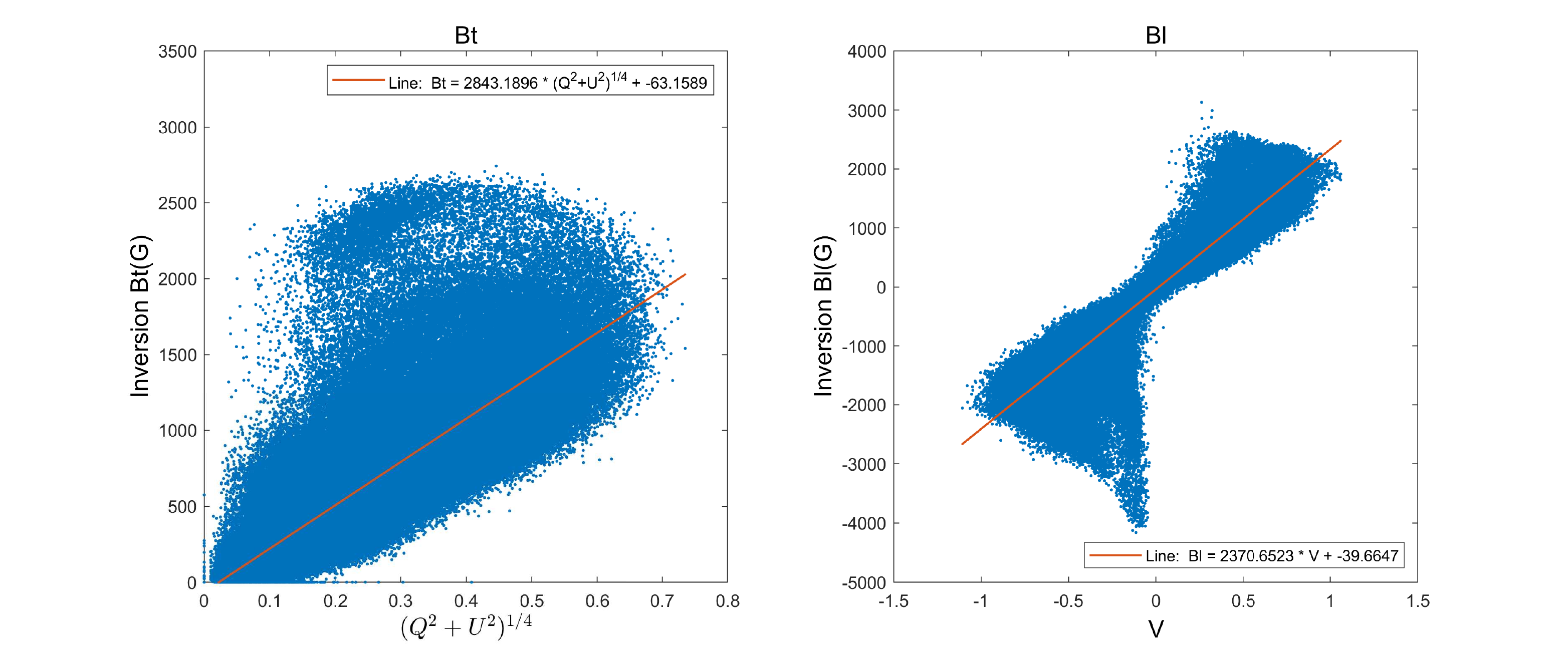}
    \caption{The scatter \textit{diagram on the left} displays the distribution relationship between the $B_t$ and $ (Q^2 + U^2)^{1/4}$. The \textit{right one} presents the relationship between $B_l$ and $V$. The \textit{red lines} are the fitted lines. The Stokes parameters $Q$, $U$, and $V$ are all normalized dividing by $I$.}
    \label{fig:8} 	
\end{figure}

We discuss the training results of $B_t$ and $B_l$, which show a relatively better accuracy compared with the models considering $\alpha$. The active region selected as test set is AR 12192, which has a complex structure, observed with \textit{Hinode}/SP at 23:41 UT on 24 October 2014. For the linear calibration, following Equations 1 and 2, the calibration coefficients are calculated by fitting the straight line between the Stokes parameters and the inversion results from \textit{Hinode}/SP using the least square method. The results of the linear calibration are shown in Figure~\ref{fig:8}.

\begin{figure}[!htbp]
	\centering
	\begin{minipage}{\textwidth}
		\centering
		\includegraphics[width=\textwidth]{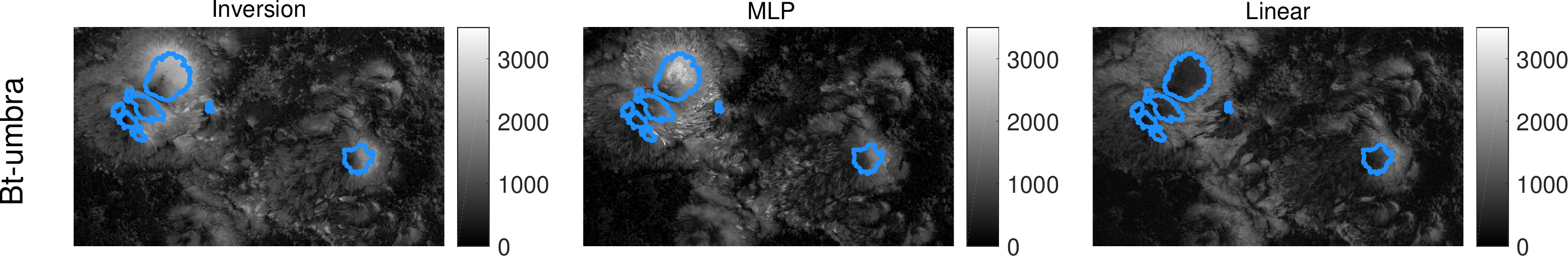}
	\end{minipage}

	\begin{minipage}{\textwidth}
		\centering
		\includegraphics[width=\textwidth]{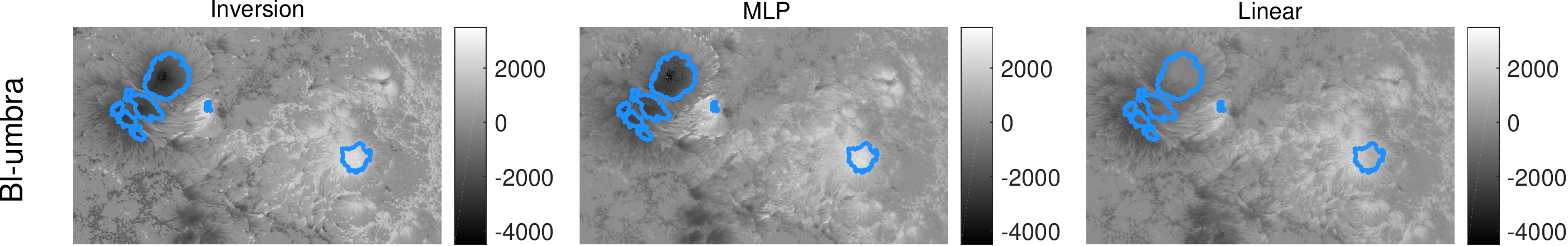}
	\end{minipage}	
	
    \caption{The results of the inversion, MagMLP, and linear calibration on the transverse field and longitudinal field. \textit{Top} is for the transverse field, \textit{bottom} for the longitudinal field.}
    \label{fig:9}
\end{figure}

Figure~\ref{fig:9} shows the results of MagMLP, linear calibration, and Stokes inversion. The umbra of the sunspot was detected using morphological reconstruction and adaptive region growing techniques \citep{Yu2014Automated}, as shown by the closed blue curves both for $B'_t$ and for $B'_l$. On the umbra region, the results of MagMLP are very close to the inversion results, while the linear calibration shows an apparent magnetic saturation effect.

\begin{figure}[!htbp]
	\centering
    \includegraphics[width=0.45\textwidth]{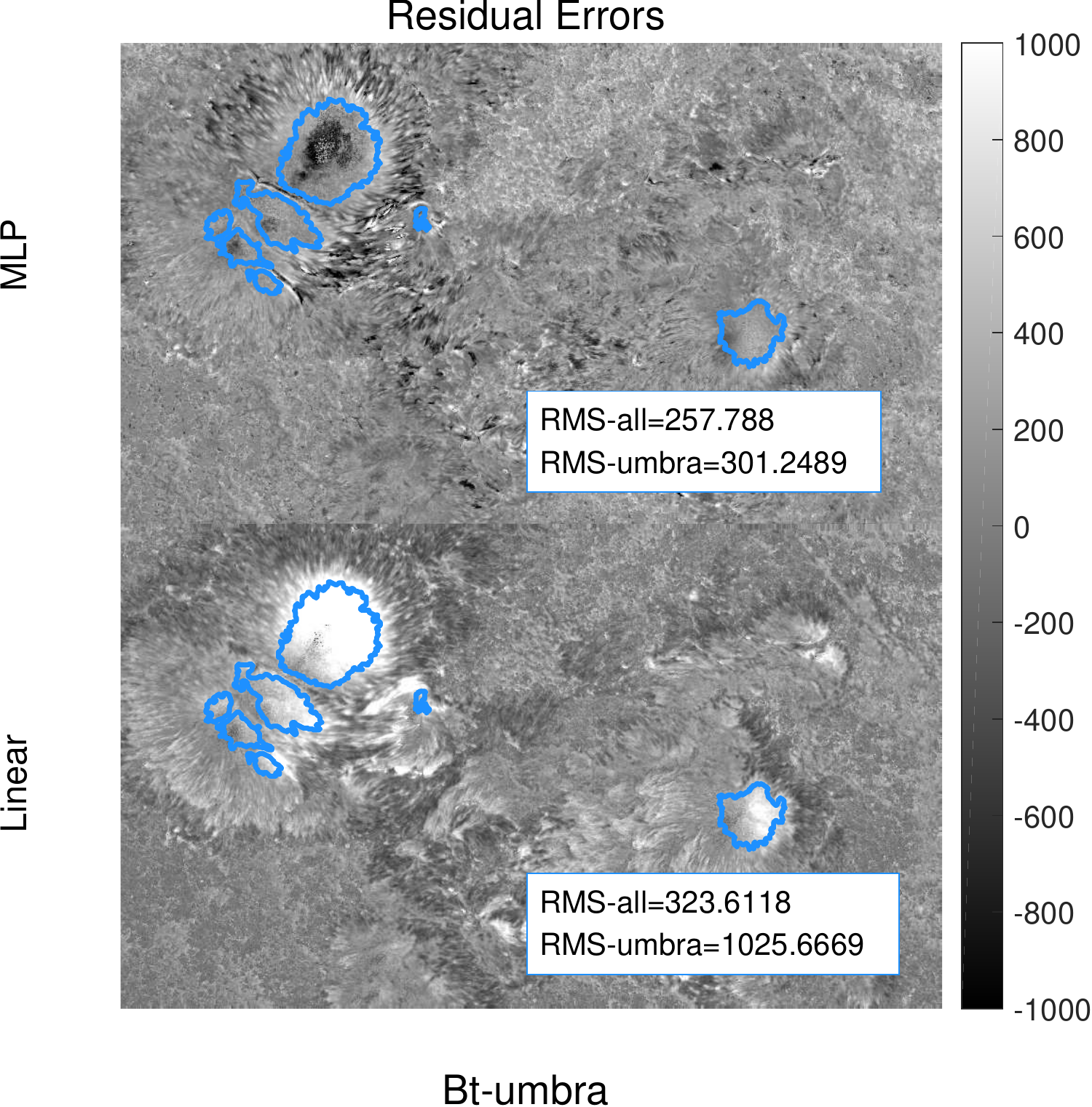}
    \qquad
    \includegraphics[width=0.4\textwidth]{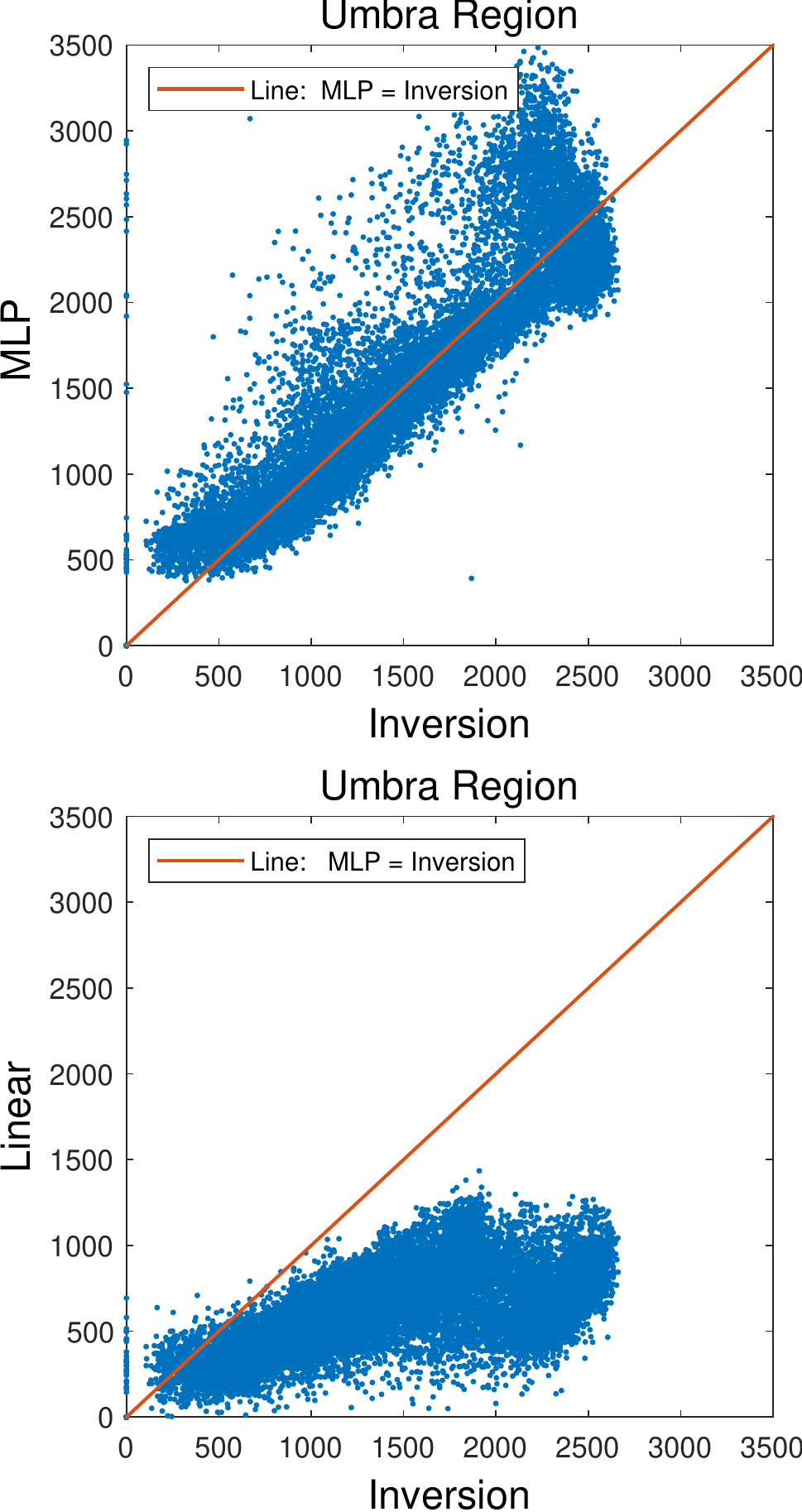}
    \caption{The results of residual errors for the transverse field. The \textit{left column} shows the residual-error outcome of MLP network models and inversion data (\textit{top}), and linear calibration results with inversion data (\textit{bottom}). The \textit{upper right panel} presents the scatter plot of the residual-error outcome of MLP network models and inversion data, the \textit{lower right panel} displays the scatter plot of the residual-error outcome of linear calibration results and inversion data.}
    \label{fig:10}
\end{figure}

The difference of results between the MLP method and the linear calibration can be inferred from the residual maps. The results of residual errors for the transverse field are displayed in Figure~\ref{fig:10}. The lower left panel of Figure~\ref{fig:10} shows the residual errors between the linear calibration results and inversion data in which the closed blue curves circle the regions with a value greater than 1000 G, even exceeding 2000 G at some pixels. This is because with the increase of magnetic field strength, the value of the corresponding Stokes parameters no longer increase in strong magnetic field regions, where the weak-field assumption is no longer applicable, that is, we have the magnetic saturation effect. The residual errors between MagMLP and inversion data are mostly within 300 G in the umbra region, and only a few are larger. Overall, the residuals of MagMLP from the inversion results are an order of magnitude lower than the residuals of the linear calibration from the inversion results. Furthermore, the RMS of the residuals are analyzed and compared. The RMS of the residual errors for MagMLP is about 257 G over the whole active region, while for the linear calibration is about 323 G, which is about 66 G larger than MagMLP. The RMS of the residual errors in the umbra region for MagMLP is about 301 G, which is about 724 G less than that for the linear calibration.

The right two panels of Figure~\ref{fig:10} shows the scatter diagrams of residual errors in the magnetic saturation region inside the blue curves. The results of the MagMLP and the linear calibration are displayed in the vertical axis, while those of the inversion in the abscissa. The results for MagMLP are closely distributed evenly above and below the line $Y = T$ except for a few points on the top right, while the results for the linear calibration deviate a lot from $Y = T$ and have a large dispersion.

\begin{figure}[!htbp]
	\centering
	\includegraphics[width=0.45\textwidth]{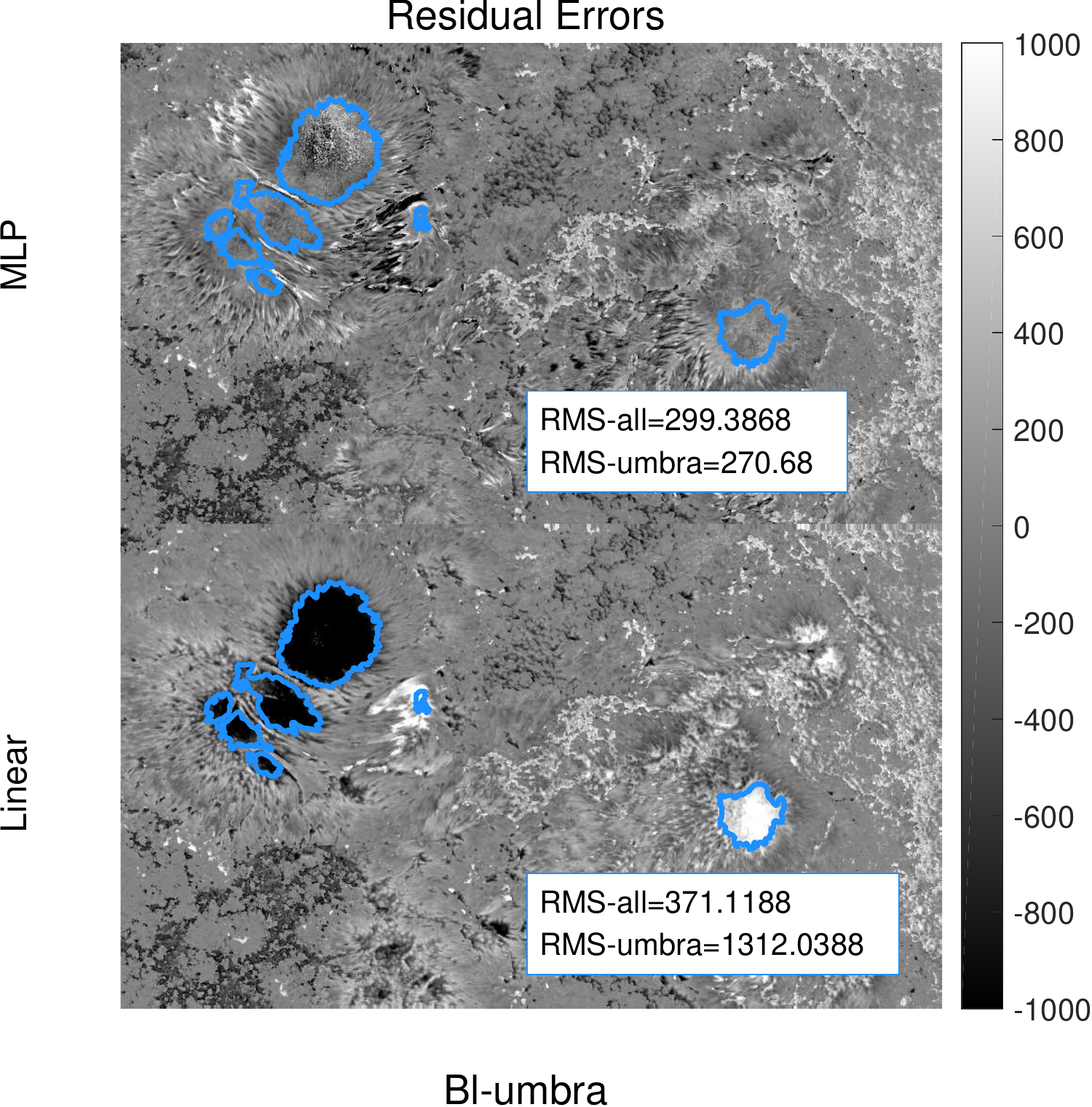}
    \qquad
	\includegraphics[width=0.4\textwidth]{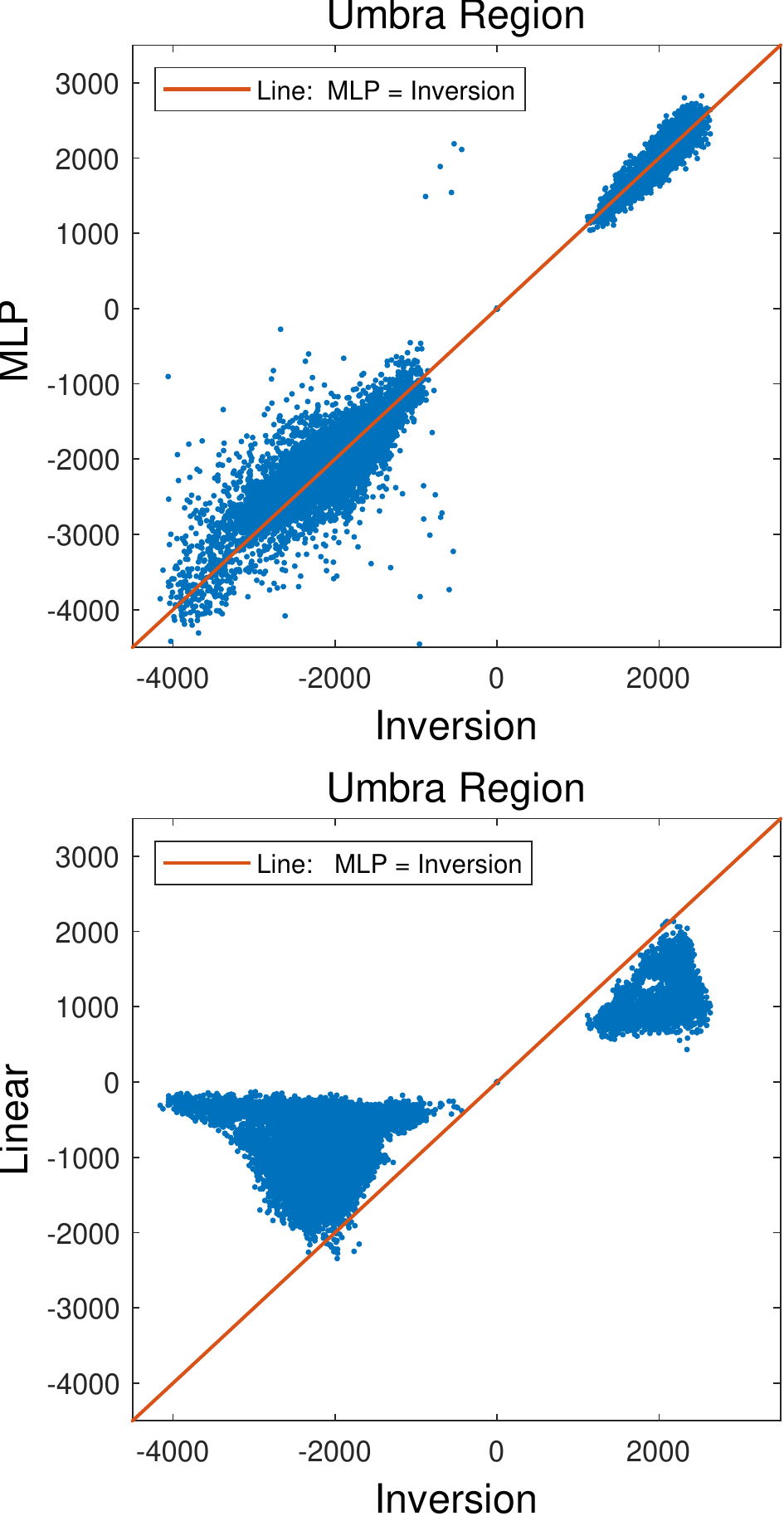}
    \caption{The results of residual errors for the longitudinal field. The \textit{left columns} shows the residual-error outcome of MLP network models and inversion data (\textit{top}), and linear calibration results with inversion data (\textit{bottom}). The \textit{upper right panel} presentes the scatter plot of residual-error outcome of MLP network models and inversion data, the \textit{lower right panel} displays the scatter plot of residual-error outcome of linear calibration results and inversion data.}
    \label{fig:11}
\end{figure}

Similar analysis for the longitudinal field is plotted in Figure~\ref{fig:11}. The lower left panel of Figure~\ref{fig:11} shows the residual errors between the linear calibration results and inversion data in which the closed blue curves circle both bright and dark regions. The results of the linear calibration have a distinct magnetic saturation effect in the umbra region and the corresponding systematic error reaches more than 1000 G in most areas of both bright and dark regions, and even exceeds 2000 G in some pixels. However, the results of MagMLP in these places are very close to the inversion results, and the systematic errors are basically within 300 G. The RMS of residual errors for the MagMLP is about 299 G over the whole active region, while for the linear calibration it is about 371 G, which is about 72 G greater than MagMLP. The RMS of residual errors in the umbra region for MagMLP is about 270 G which is about 1042 G less than that for the linear calibration. In the right panels of Figure~\ref{fig:11}, the results for MagMLP are closely distributed evenly above and below the line $Y = T$ except for a few points, while the results for the linear calibration deviate a lot from $Y = T$ and have a large dispersion in both the positive and negative polarities.

\subsection{``Bright Spots'' and ``Dark Spots'' Analysis Done on the Inclination Angle}

\begin{figure}[!htbp]
	\centering
	\begin{minipage}{\textwidth}
		\centering
		\includegraphics[width=0.42\textwidth]{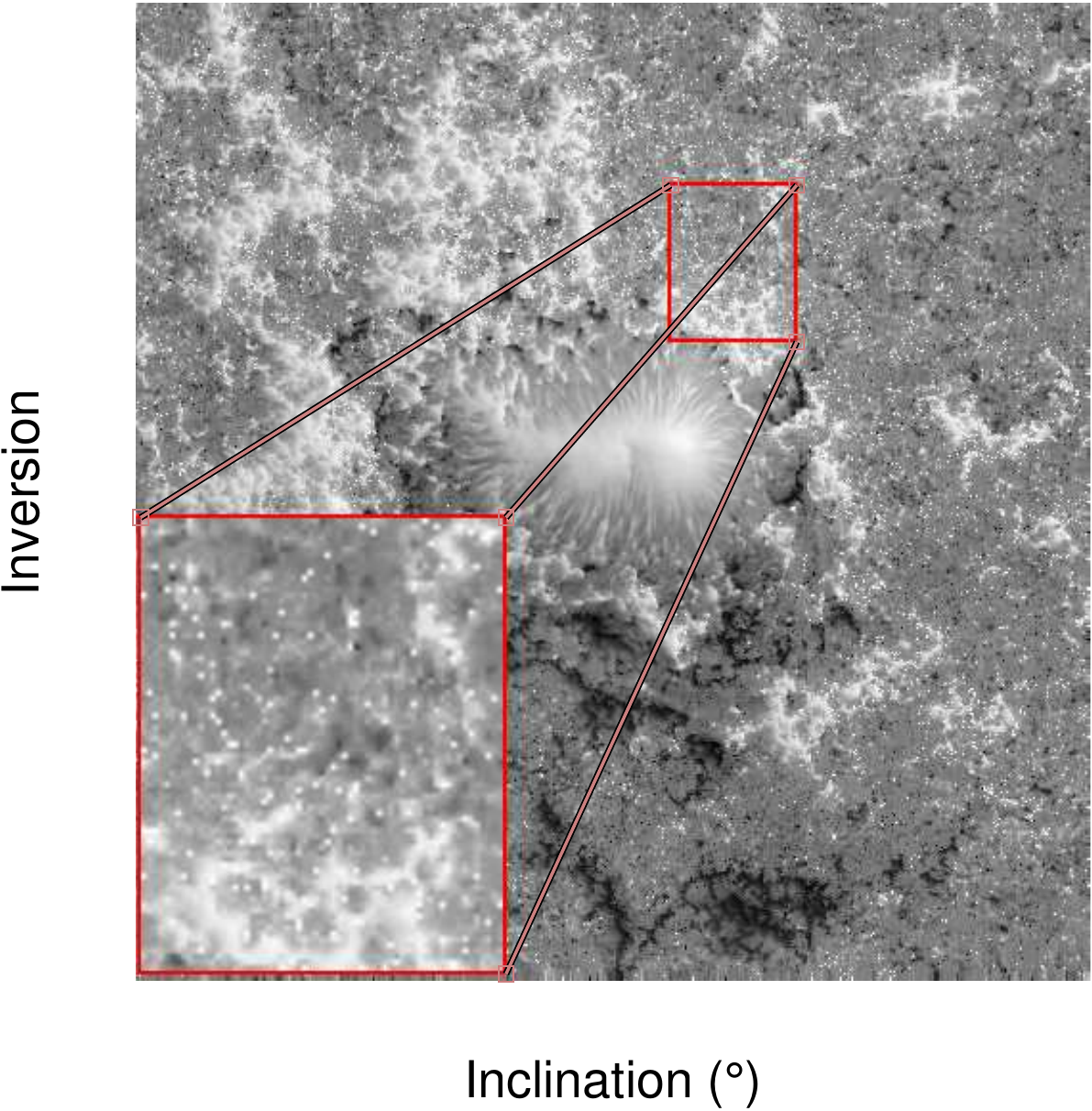}
        \qquad
		\includegraphics[width=0.42\textwidth]{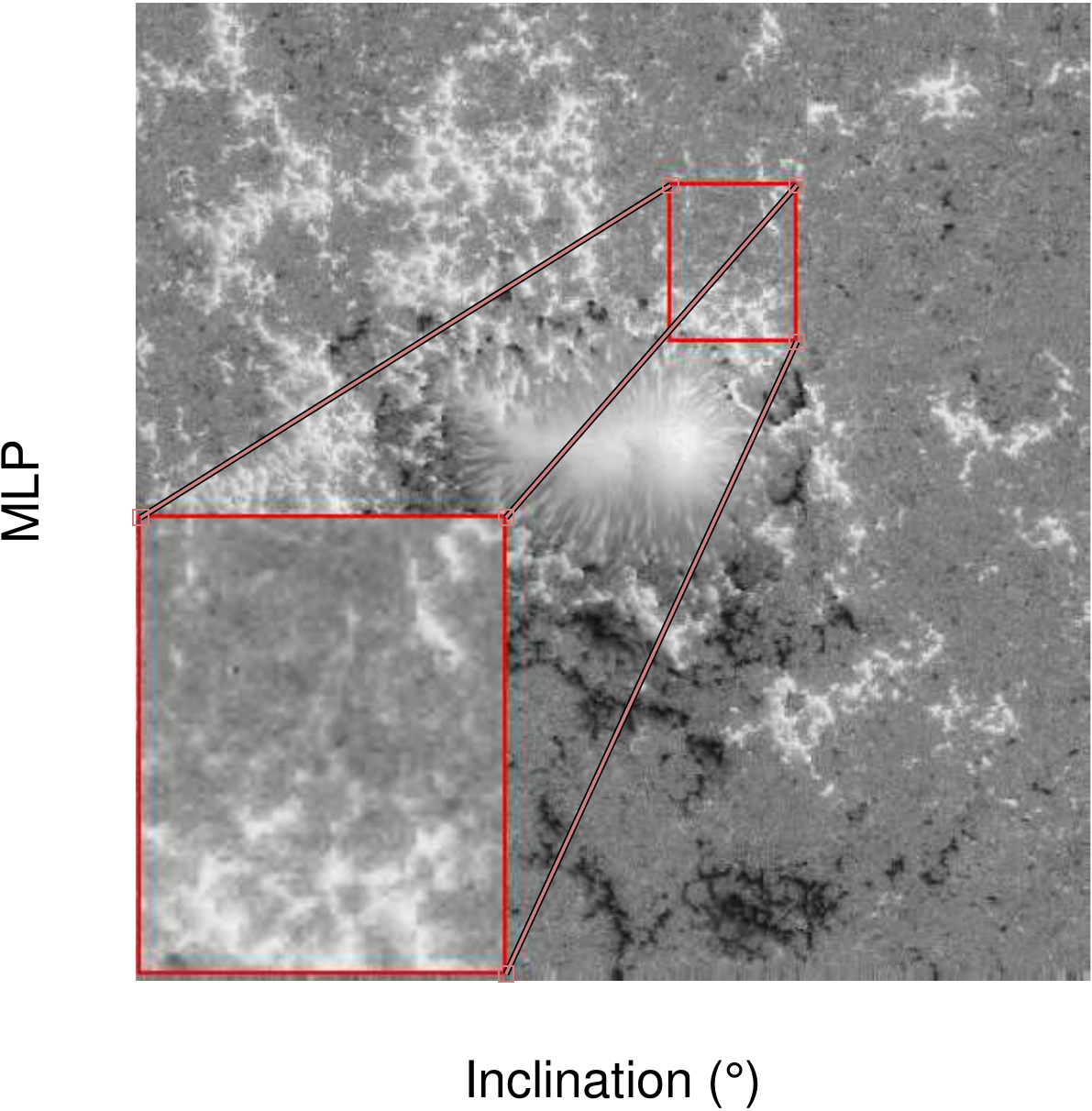}
	\end{minipage}	
	
	\begin{minipage}{\textwidth}
		\centering
		\includegraphics[width=0.5\textwidth]{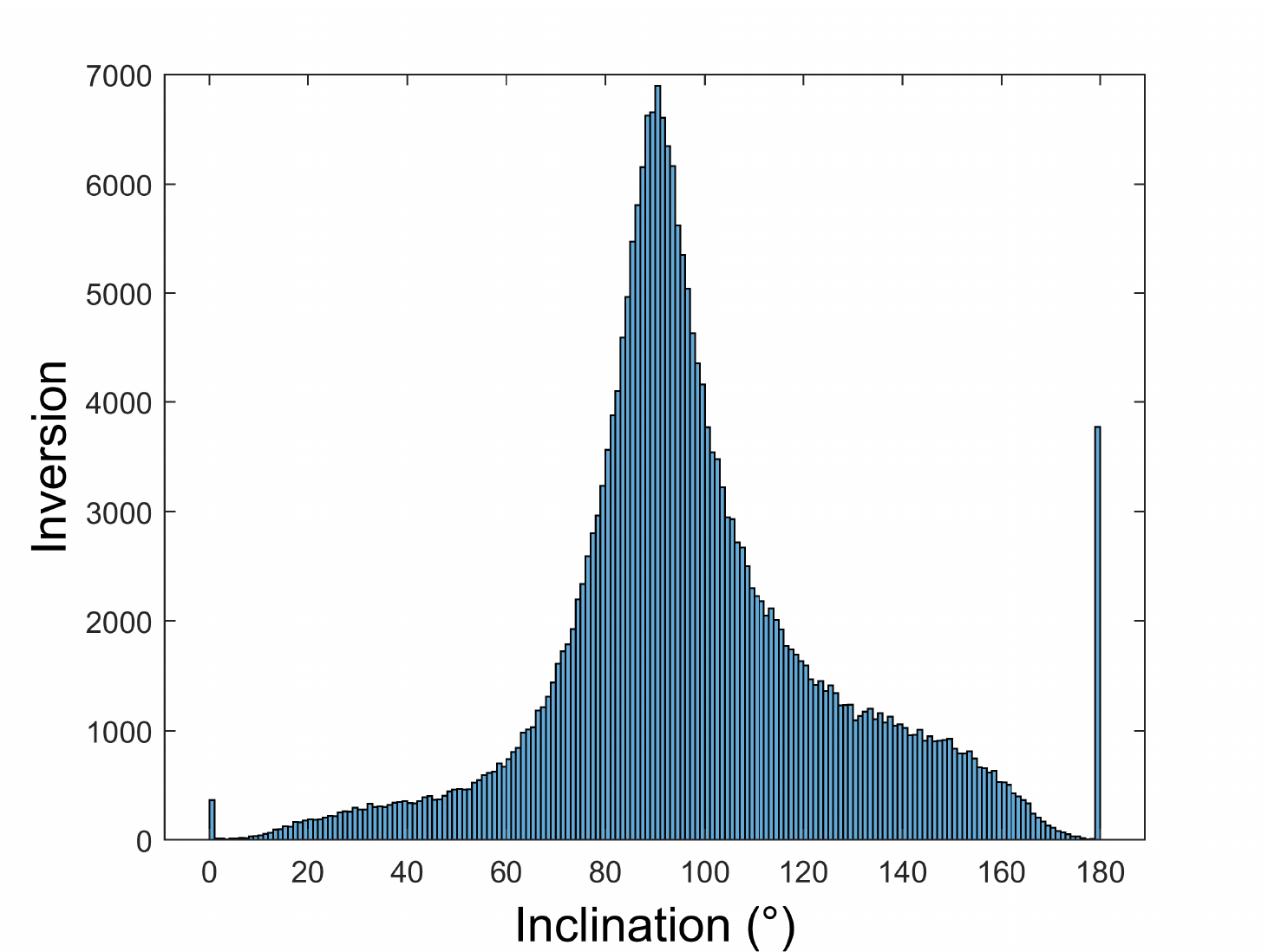}
        \hspace*{-0.025\textwidth}
		\includegraphics[width=0.5\textwidth]{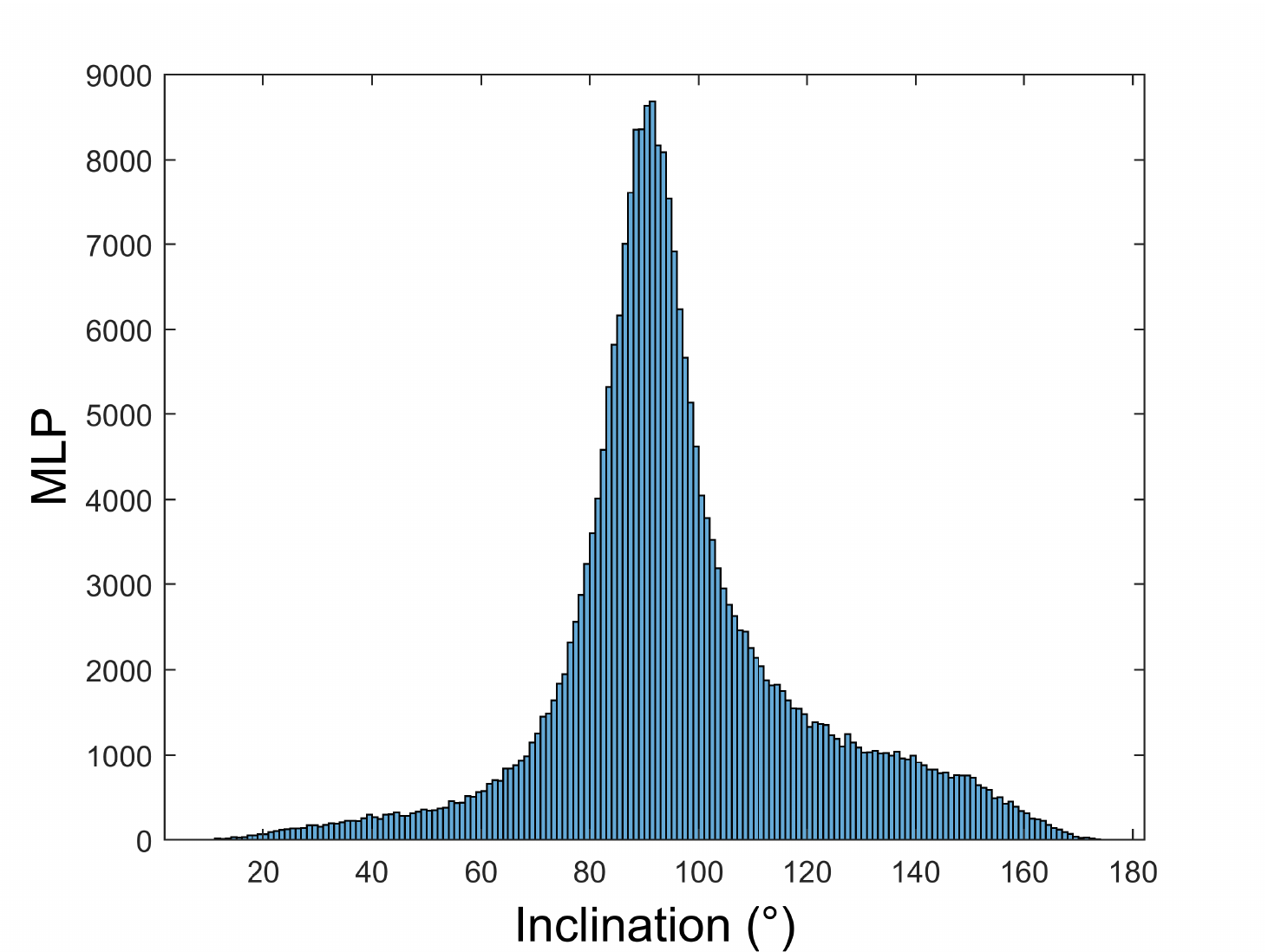}
	\end{minipage}
	
    \caption{The comparison of the inclination angle between the inversion and MagMLP. The \textit{upper left panel} is the map of results for the inversion and the \textit{lower left panel} is its histogram, the \textit{upper right panel} is for MagMLP and the \textit{lower right panel} is its histogram.}
    \label{fig:12}
\end{figure}

In the inclination angle data of \textit{Hinode}/SP using the Stokes inversion, we notice that there are many ``bright spots'' and ``dark spots'', which have a large contrast with the surrounding regions. Their values are 180 and 0 degrees as displayed on the upper left panel in Figure~\ref{fig:12}. Let us take the ``bright spots'' as an example to illustrate. The active region selected as test set is AR 12740, observed with \textit{Hinode}/SP at 06:00UT on 8 May 2019. Many ``bright spots'' can be seen in the results of the inversion data zoomed in on the red rectangular box. At these pixels the Stokes inversion is not reliable and does not obtain good results. While in the results of MagMLP, these ``bright spots'' vanish and the whole region in the red rectangular box (magnified) looks smooth without a lot of obvious bright points as shown on the upper right panel in Figure~\ref{fig:12}. It means that the MagMLP can give relatively reasonable and reliable results for these points. The histograms (bottom row in Figure~\ref{fig:12}) show that the MagMLP results have no 180 degree points as shown on the lower right, while the inversion includes about 3900 points with the value of 180 degrees displayed on the lower left of Figure~\ref{fig:12}. A similar phenomenon appears on the ``dark spots''. This results suggests that the MagMLP works well on these points. Generally, the observed polarization profiles are dominated by noise with insufficient signals in these pixels, where the Stokes inversion fails, while the MagMLP has the capability to obtain more reasonable results.

\section{Conclusion and Future Work}
\label{sec:conc}

The magnetic field calibration on a single wavelength is essentially a multiple regression problem. Multilayer neural networks could effectively perform complex multiple regression and flexibly set the number of input variables, which is a novel machine-learning method and worth applying to the magnetic field calibration. Based on multilayer fully connected neural networks, the magnetic field calibration for filter-based magnetographs is done through a hierarchical mapping to build the network model MagMLP using data from \textit{Hinode}/SP to simulate single-wavelength observations for model training.

As we know, the linear calibration for filter-based magnetographs has a significant magnetic saturation effect in the regions with strong magnetic field. We compared the MagMLP with the linear calibration method, pointing out that the MagMLP shows much better performance on solving the magnetic saturation effect. From the analysis of an active region, the results of the linear calibration show evident magnetic saturation effect in the umbra region and the corresponding systemic error reaches values greater than 1000 G in most areas, or even exceeds 2000 G at some pixels. However, the results of MagMLP are very close to the inversion data, and the systematic errors are basically within 300 G.

The training results show that the linear fitting coefficient of the transverse field to the target reaches above 0.91, and that of the longitudinal field is above 0.98. The generalization ability of the models is good, as the LFCs of all the models are above 0.9 on the test subsets. In addition, the Doppler velocity field and the filling factor have influence to different extents on the results. In a word, the simpler the relationship between input and output parameters is, the better the results are.

In addition, we find that there are many ``bright spots'' and ``dark spots'' on the inclination angle images from the inversion results of \textit{Hinode}/SP with values of 180 and 0 degrees, respectively, where the inversion is not reliable and does not produce a good result but the MagMLP could handle these points well.

It can be seen that the MagMLP can converge well and perform with good robustness and generalization ability. It is important to emphasize that this method is only a preliminary attempt as an auxiliary mean at present. In order to obtain the magnetic field strength and the inclination angle, which is the output of the MagMLP, we can scan the wavelength with a tunable filter to get the Stokes $I$, $Q$, $U$, and $V$ profiles and carry out the Stokes inversion. Regarding FMG, we have also designed an on-orbit calibration mode.

For magnetic field strength, a pixel and its neighboring pixels must not be isolated, but correlated with each other, which has been neglected in the preliminary attempt done in this article. Next, we will consider the relationship between adjacent regions of the magnetic field and its two-dimensional structure, and use the Convolutional Neural Network to study the calibration of the magnetic field in order to obtain better accuracy.

\acknowledgements

We are grateful to the Astronomical Big Data Joint Research Center, co-founded by the National Astronomical Observatories, the Chinese Academy of Sciences, and the Alibaba Cloud. We are thankful to Prof. Song Feng from Kunming University of Science and Technology for providing the code to extract the umbra of solar active region. This project has received funding from the Strategic Priority Research Program on Space Science, the Chinese Academy of Sciences under No. XDA15320300, XDA15320302, XDA15052200, XDA15010800, the National Natural Science Foundation of China (NSFC) under No.11873027, 11773072, 11427803, 11427901, 11773040, 11573012, 11833010, 11973056, 11873062, 11703042, and Beijing Municipal Science and Technology under No. Z181100002918004.


\end{document}